\documentclass[twocolumn,aps,prl,superscriptaddress]{revtex4-2}

\usepackage{color,amsmath,amssymb,epsfig,graphicx}
\usepackage{bm}
\usepackage{mathptmx, courier, pifont}
\usepackage[scaled=0.92]{helvet}
\usepackage[T1]{fontenc}
\usepackage{textcomp}
\usepackage[colorlinks=true,linkcolor=blue,filecolor=blue,urlcolor=blue,citecolor=blue]{hyperref}

\usepackage{multirow}
\usepackage{longtable}
\usepackage[utf8]{inputenc}

\makeatletter
\def\@seccntformat#1{\csname the#1\endcsname.\quad}
\makeatother
\begin{document}

\title{ Microscopic investigation of  $\gamma~$ vibrational band structures in odd-mass nuclei }

\author{Uzma Jahangir}
\affiliation{Department of  Physics, Islamic University of Science and Technology, Awantipora, 192 122, India}
\author{S.~P.~Rouoof} \email{sprouoofphysics27@gmail.com}
\affiliation{Department of  Physics, Islamic University of Science and Technology, Awantipora, 192 122, India}
\author{S.~Jehangir}
\affiliation{Department of Physics, Government Degree College Kulgam, Jammu and Kashmir Higher Education Department, 192 231, India}
\author{G.~H.~Bhat}
\affiliation{Department of Physics, Government Degree College Shopian, Jammu and Kashmir Higher Education Department, 192 303, India}
\author{J.~A.~Sheikh} \email{sjaphysics@gmail.com}
\affiliation{Department of Physics, University of Kashmir, Srinagar, 190 006, India}
\author{N.~A.~Rather}
\affiliation{Department of  Physics, Islamic University of Science and Technology, Awantipora, 192 122, India}

\begin{abstract}

  A systematic investigation of the high-spin band structures observed in $^{103,105,107,109}$Nb and $^{103,105,107,109}$Tc nuclides
  is performed using the triaxial
  projected shell model (TPSM) approach. For $^{103,105}$Nb isotopes, four bands have been populated with the lowest three bands
  corresponding to yrast, $\gamma$ and 2$\gamma$ bands. The nature of the fourth observed band has remained unresolved as it has been
  shown from the transition intensity ratios  that this band cannot correspond to the expected 3$\gamma$ band. It is demonstrated in the present work that
  this fourth band is the second $\gamma$ band, resulting from the combination, $K=K_0-2$ with $K_0$ being the
  "$K$" value of the parent configuration. The excitation energy and other properties of this band structure are predicted for all the studied
  nuclides. 

\end{abstract}
\date{\today}
\maketitle

\section{Introduction}
\label{Sect.01}

The development of phenomenological models in nuclear physics has played a pivotal role in our understanding of excitation modes in atomic
nuclei. The collective model of Bohr and Mottelson laid the foundation of the phenomenological models in nuclear physics \cite{BMII}.
It was demonstrated using this model that nuclear collective excitation modes across the periodic table can be broadly
categorized in terms of vibrational and rotational degrees of freedom. The model considers
the quadrupole field as the elementary excitation mode responsible for the collective excitations. The collective model Hamiltonian
is expressed in terms of the quantized forms of the coordinates of quadrupole shape parameters of $\beta$ and $\gamma$,
and the three Euler angles.
It is known that many nuclear properties observed at low-excitation energies in deformed and transitional nuclei can be explained
using this simplistic view of the atomic nucleus. In particular, the low-lying excited bands observed in atomic nuclei can simply
be explained as oscillations in $\beta$ and $\gamma$ coordinates of an ellipsoidal body. $\beta$ vibration is the oscillation of nuclear
shape along the symmetry axis, preserving the axial symmetry of the system with the phonon excitation mode having $K^\pi =0^+$. The first
excited $0^+$ states observed in deformed nuclei are generally classified as corresponding to the $\beta$ vibrational mode.
$\gamma$ vibration, on the other hand, is the
oscillation of the nuclear shape perpendicular to the symmetry axis, breaking the axial symmetry with the phonon excitation
having $K^\pi = 2^+$. The first excited band with band head having $I^\pi=2^+$, observed in most of the deformed and transitional nuclei,
is categorized as the $\gamma$ vibrational band \cite{BMII}.

In recent years, the microscopic approach of triaxial projected shell model (TPSM) has provided some new insights into the
structure of the $\gamma$ bands observed in atomic nuclei. It has been shown that $\gamma$ bands are built on each intrinsic triaxial
configuration \cite{sh10,SJ18,JS21}. In general, a triaxial configuration is a superposition of several ``$K$'' states. However,
due to symmetry requirement
only particular ``$K$'' values are permitted \cite{SJ18,JS21,SH99} for a given intrinsic state. For instance, for 
the triaxial self-conjugate vacuum state, only even-values of $K$= 0,2, 4,... are allowed \cite{KY95,JS16,SJ18,JS21}. The angular momentum projection
from $K$= 0, 2, 4,... states then gives rise to yrast, $\gamma$, $2\gamma$,... band structures observed in deformed
nuclei \cite{YK00}. It is evident from this perspective that each triaxial state should have $\gamma$ bands built on it. As a matter
of fact, these bands have now been observed to be built on two-quasiparticle states in even-even \cite{PhysRevC.93.034317,JS21}, odd-odd systems
\cite{1gy6-v3sb}, and one- and three quasiparticle states in odd-mass systems \cite{WANG2009420,Chakraborty_2023,Jeh21}.

\begin{figure*}[htb]
\vspace{0cm}
\includegraphics[totalheight=14cm]{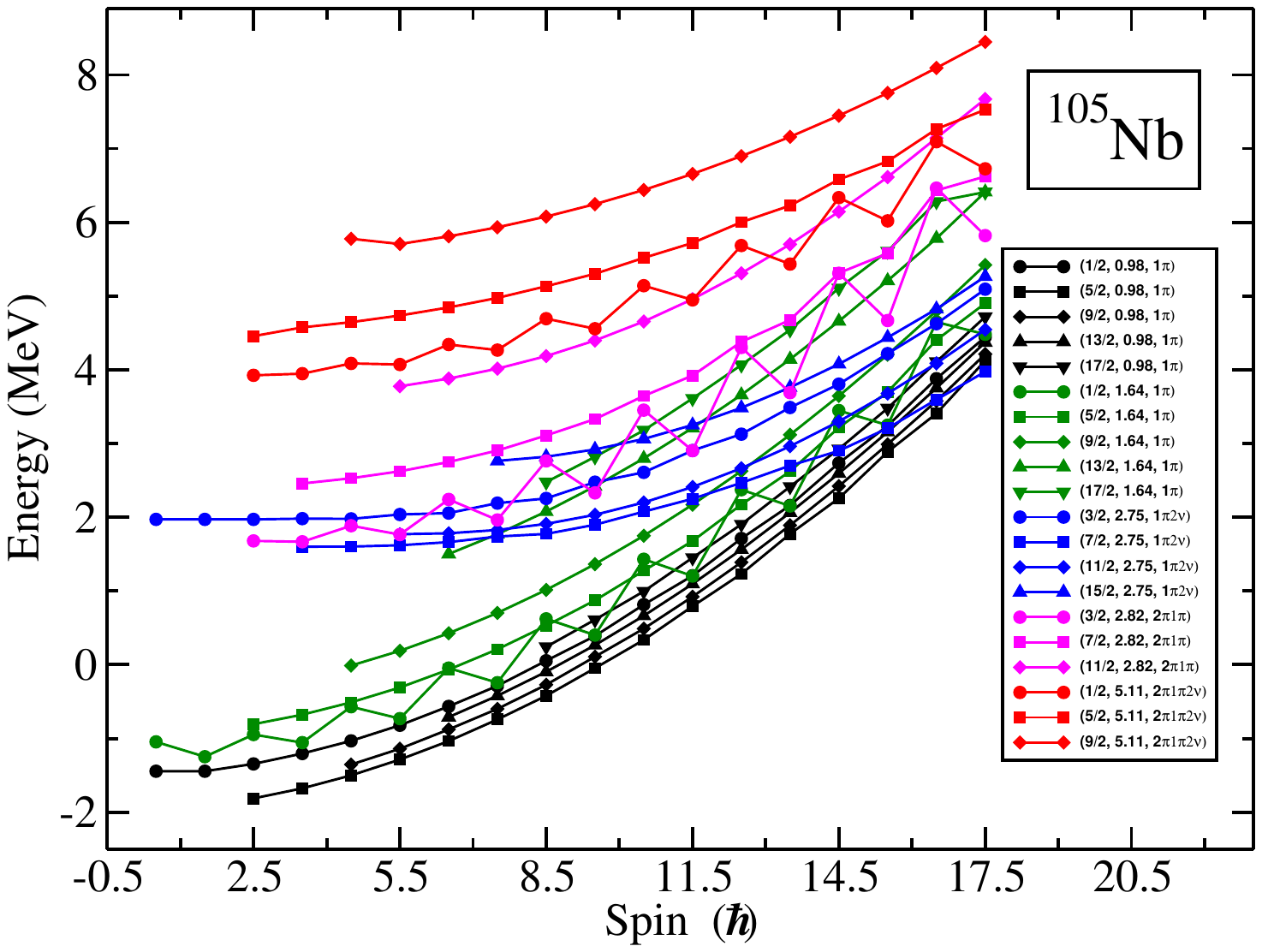}
\caption{(Color
online)  Projected energies are shown before diagonalization of the
shell model Hamiltonian for $^{105}$Nb. The bands are labelled by
three quantities : $K$-quantum number,  energy and group structure of the quasiparticle
state. For instance, $(1/2, 0.98, 1\pi)$ designates one-quasiproton
state having intrinsic energy of 0.98 MeV and K$=1/2$.  } \label{bd1}
\end{figure*}
\begin{figure*}[htb]
\vspace{0cm}
\includegraphics[totalheight=13cm]{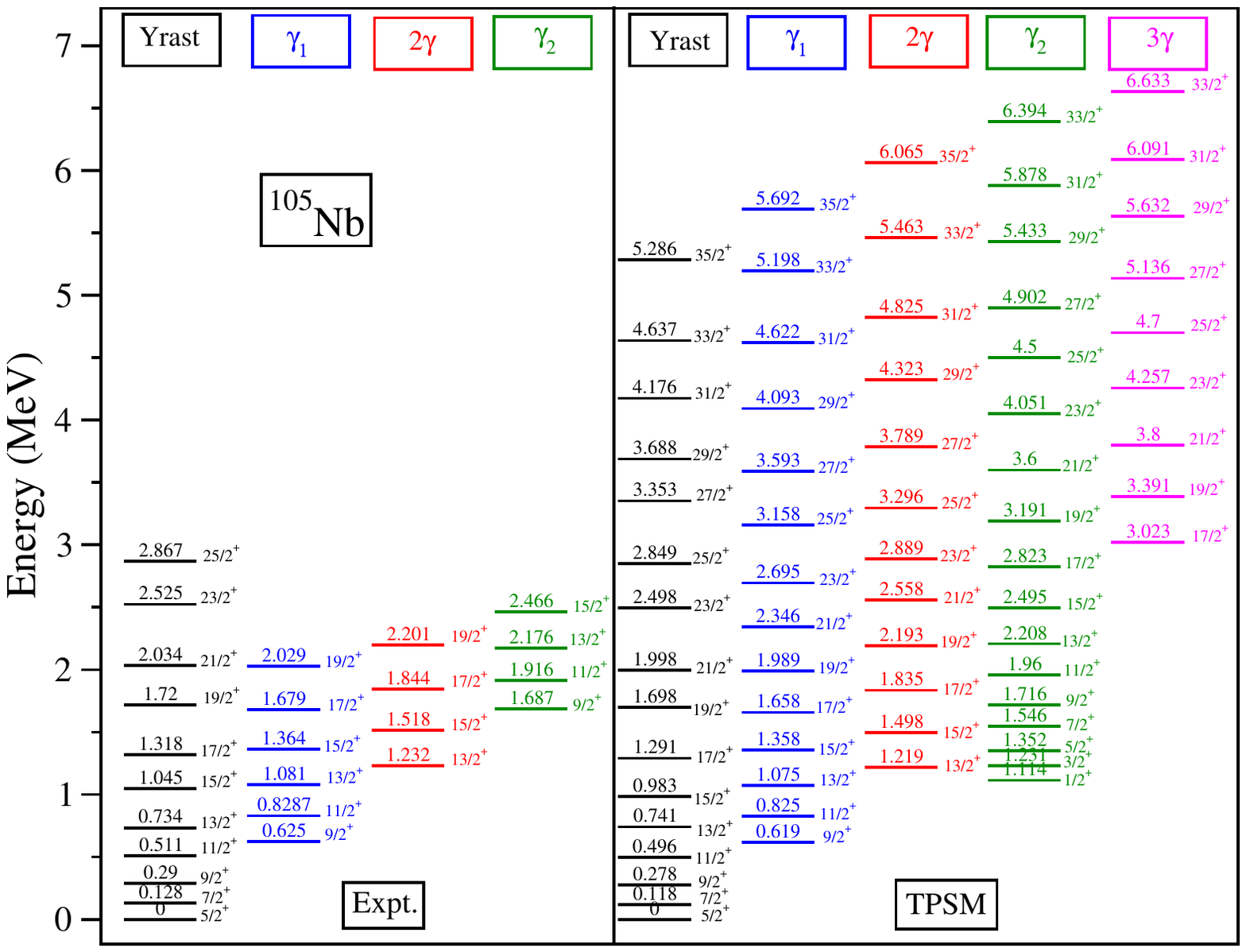} \caption{(Color
online) TPSM energies for the lowest five bands after configuration
mixing are plotted along
with the available experimental data for $^{105}$Nb isotope. Data are
taken from \cite{Li054311}.} \label{105Nb}
\end{figure*}
\begin{figure*}[htb]
\vspace{0cm}
\begin{center}
\includegraphics[totalheight=13cm]{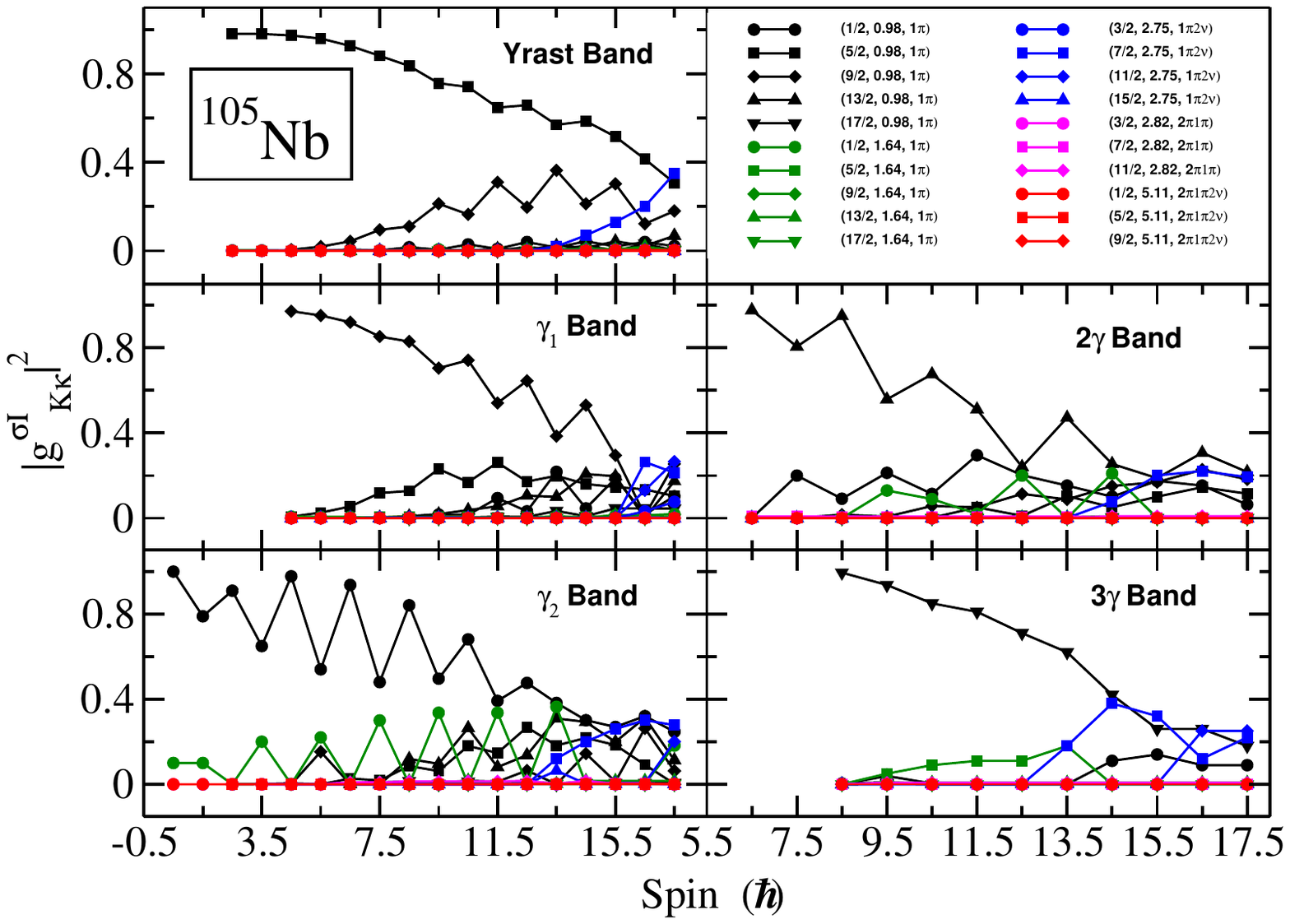}
\caption{(Color
online)  Probabilities of various projected $K$ configurations in the orthonormal basis of $^{105}$Nb. 
    The curves are labelled by three quantities: $K$ quantum number, energy of the quasiparticle state and quasiparticle character. For instance, [1/2, 0.98, 1$\pi$] designates one quasiproton state with $K$= 1/2 having intrinsic energy of 0.98 MeV. 
  } \label{wavefun}
\end{center}
\end{figure*}
\begin{figure}[htb]
\vspace{0cm}
\begin{center}
\includegraphics[totalheight=12cm]{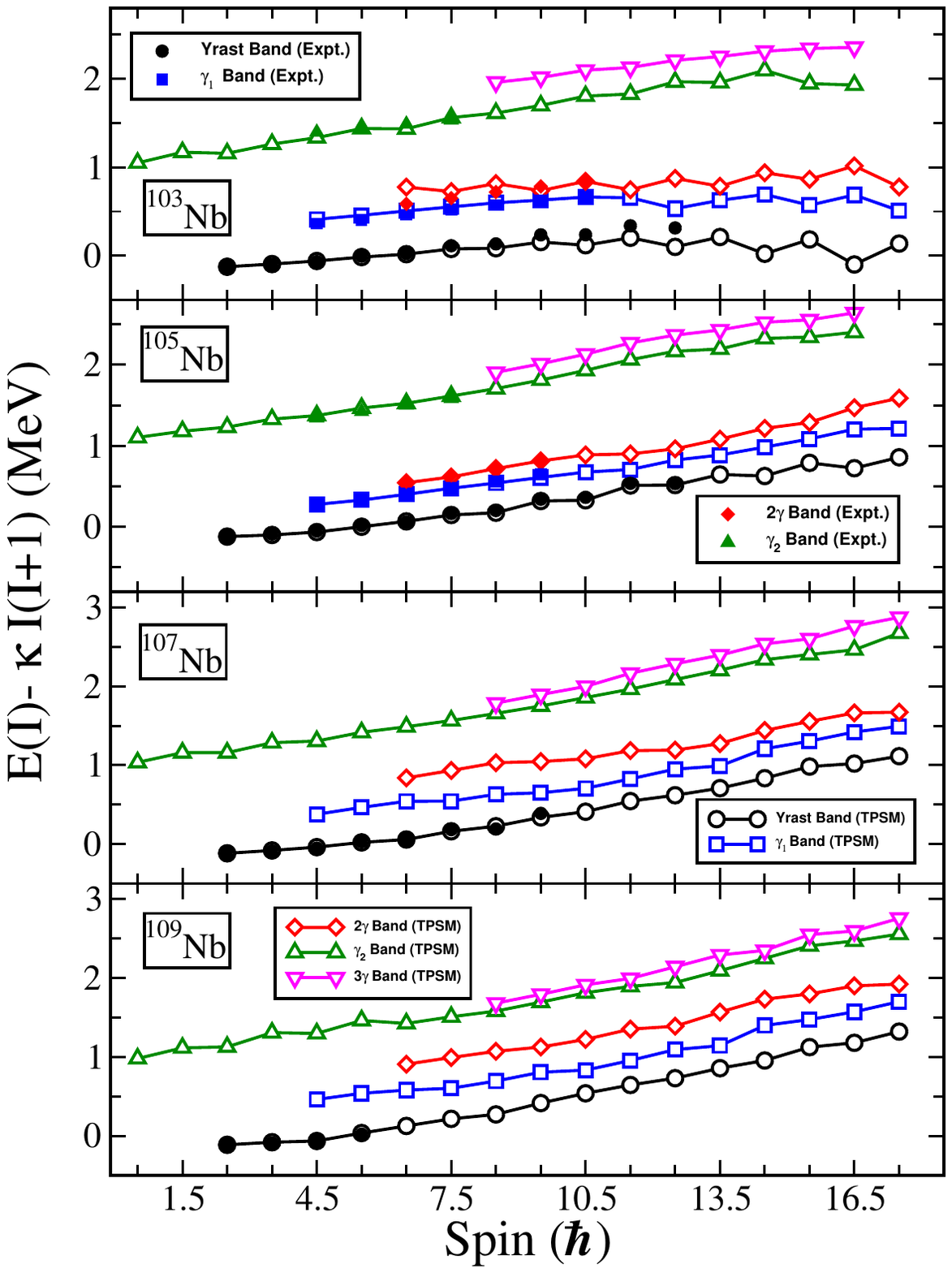}
\caption{(Color
online) TPSM and experimental energies of the Yrast, $\gamma_1$, $2\gamma$, $\gamma_2$ and  $3\gamma$ bands of $^{103,105,107,109}$Nb isotopes. The scaling factor,  $\kappa=32.32A^{-5/3}$.
  } \label{eng_core_Nb}
\end{center}
\end{figure}
\begin{figure}[htb]
\vspace{0cm}
\begin{center}
\includegraphics[totalheight=12cm]{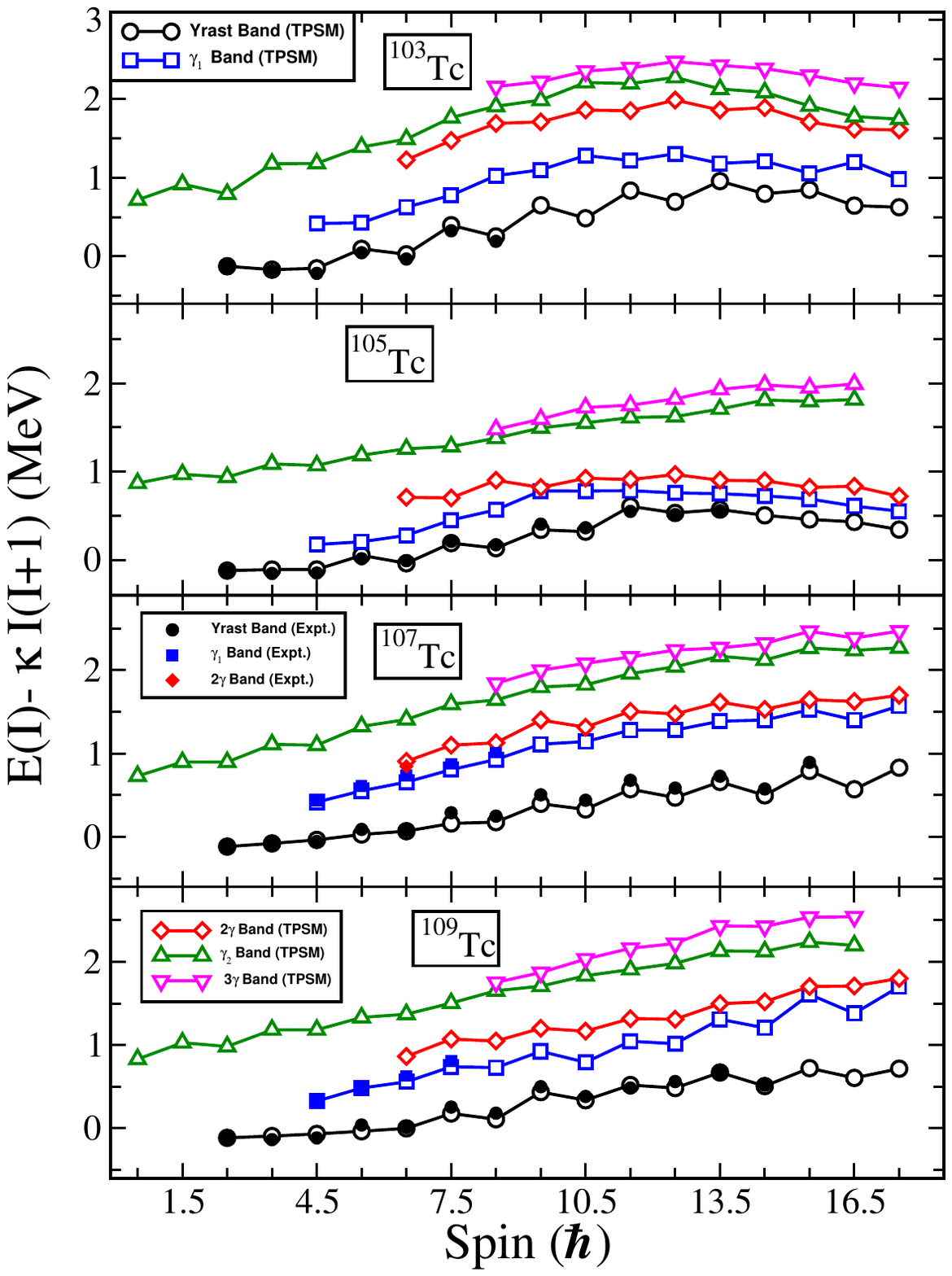}
\caption{(Color
online) TPSM and experimental energies of the Yrast, $\gamma_1$, $2\gamma$, $\gamma_2$ and $3\gamma$ bands of $^{103,105,107,109}$Tc isotopes. The scaling factor,  $\kappa=32.32A^{-5/3}$.
  } \label{eng_core_Tc}
\end{center}
\end{figure}
\begin{figure}[htb]
\vspace{0cm}
\begin{center}
\includegraphics[totalheight=12cm]{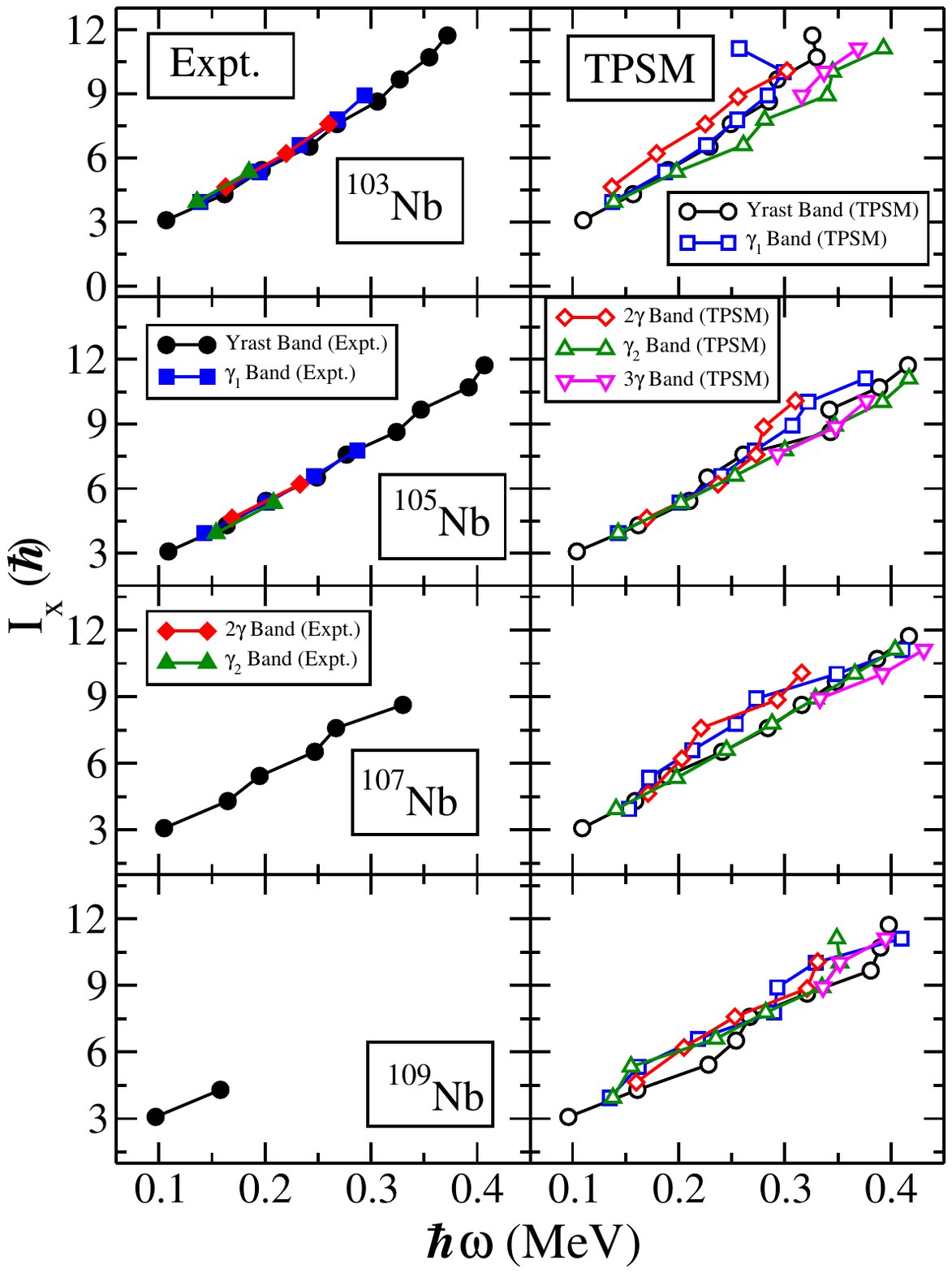} \caption{(Color
online) TPSM angular momentum $I_x=\sqrt{I(I+1)-K^2}$ versus the rotational frequency $\hbar \omega=\frac{\Delta E}{\sqrt{(I+1.5)^2-K^2}-\sqrt{(I+0.5)^2-K^2}}$ for the four bands are plotted along
with the available experimental data for $^{103,105,107,109}$Nb isotopes. } \label{Nb_align}
\end{center}
\end{figure}
\begin{figure}[htb]
\vspace{0cm}
\begin{center}
\includegraphics[totalheight=12cm]{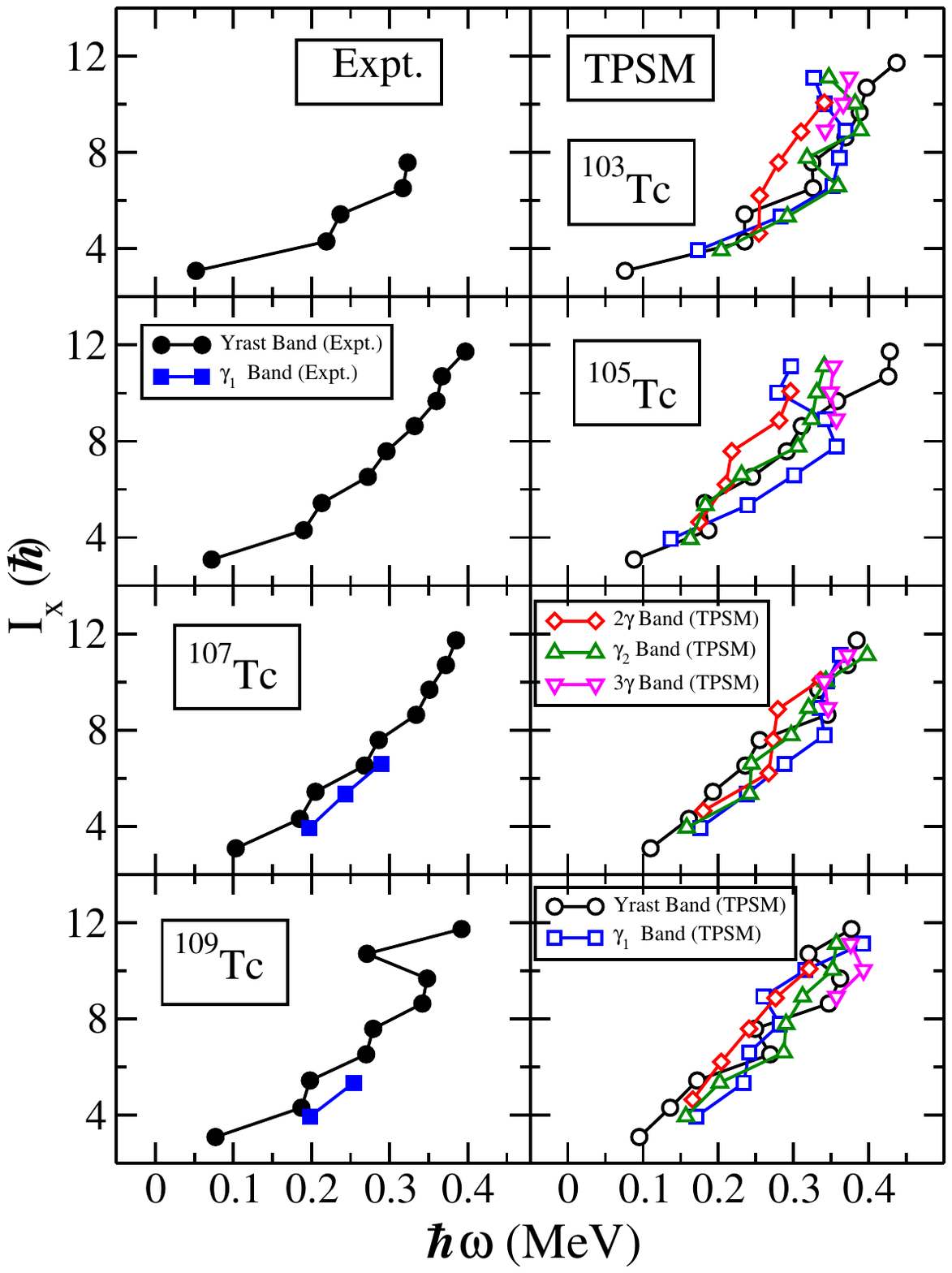} \caption{(Color
online) TPSM angular momentum $I_x=\sqrt{I(I+1)-K^2}$ versus the rotational frequency $\hbar \omega=\frac{\Delta E}{\sqrt{(I+1.5)^2-K^2}-\sqrt{(I+0.5)^2-K^2}}$ for the four bands are plotted along
with the available experimental data for $^{103,105,107,109}$Tc isotopes.} \label{Tc_align}
\end{center}
\end{figure}
\begin{figure*}[htb]
\vspace{0cm}
\begin{center}
\includegraphics[totalheight=13cm]{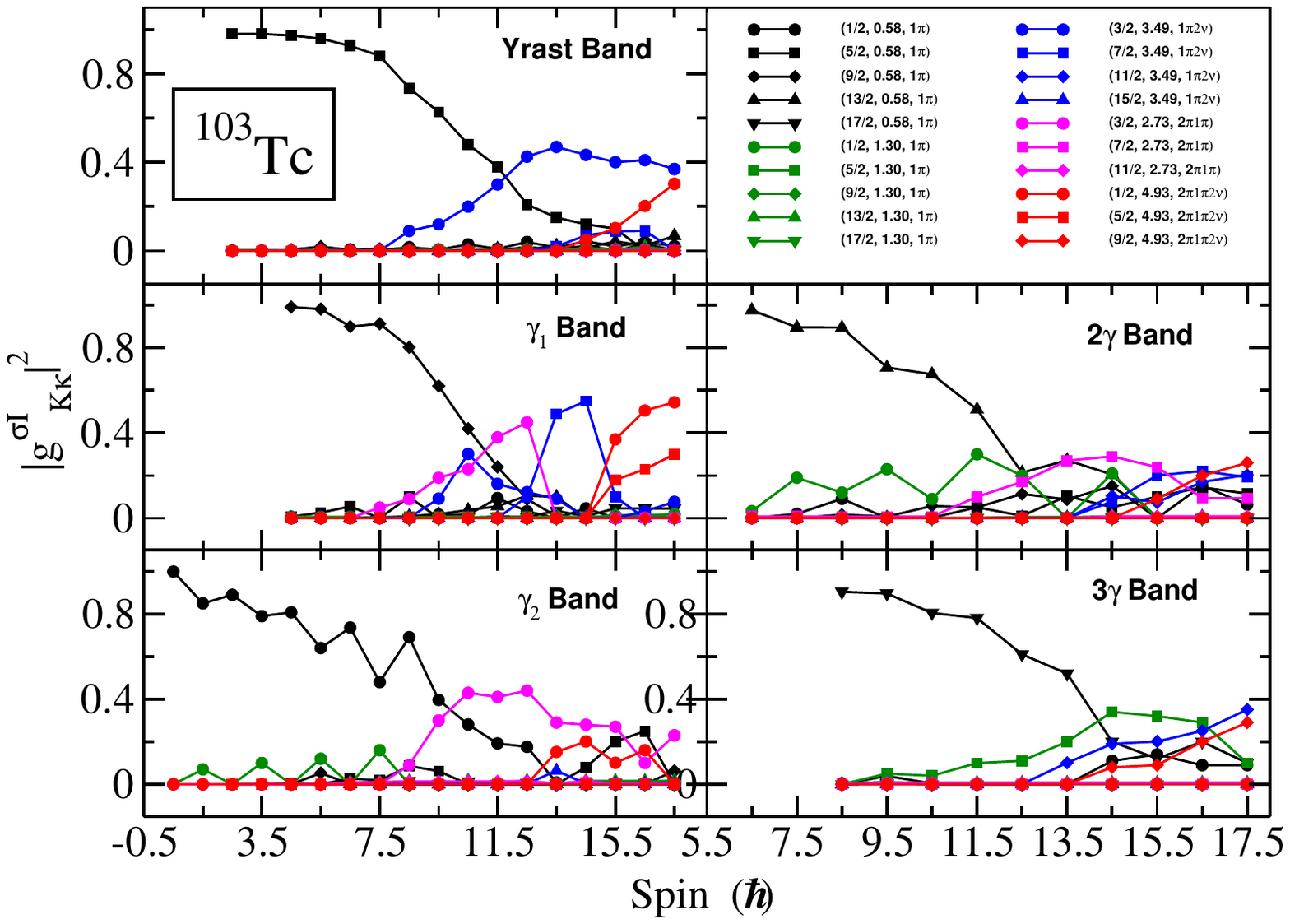}
\caption{(Color
online)  Probabilities of various projected $K$ configurations in the orthonormal basis of $^{103}$Tc. The curve labels have the same meaning as in Fig. \ref{wavefun}. 
  } \label{wavefun103Tc}
\end{center}
\end{figure*}
\begin{figure}[htb]
\vspace{0cm}
\begin{center}
\includegraphics[totalheight=12cm]{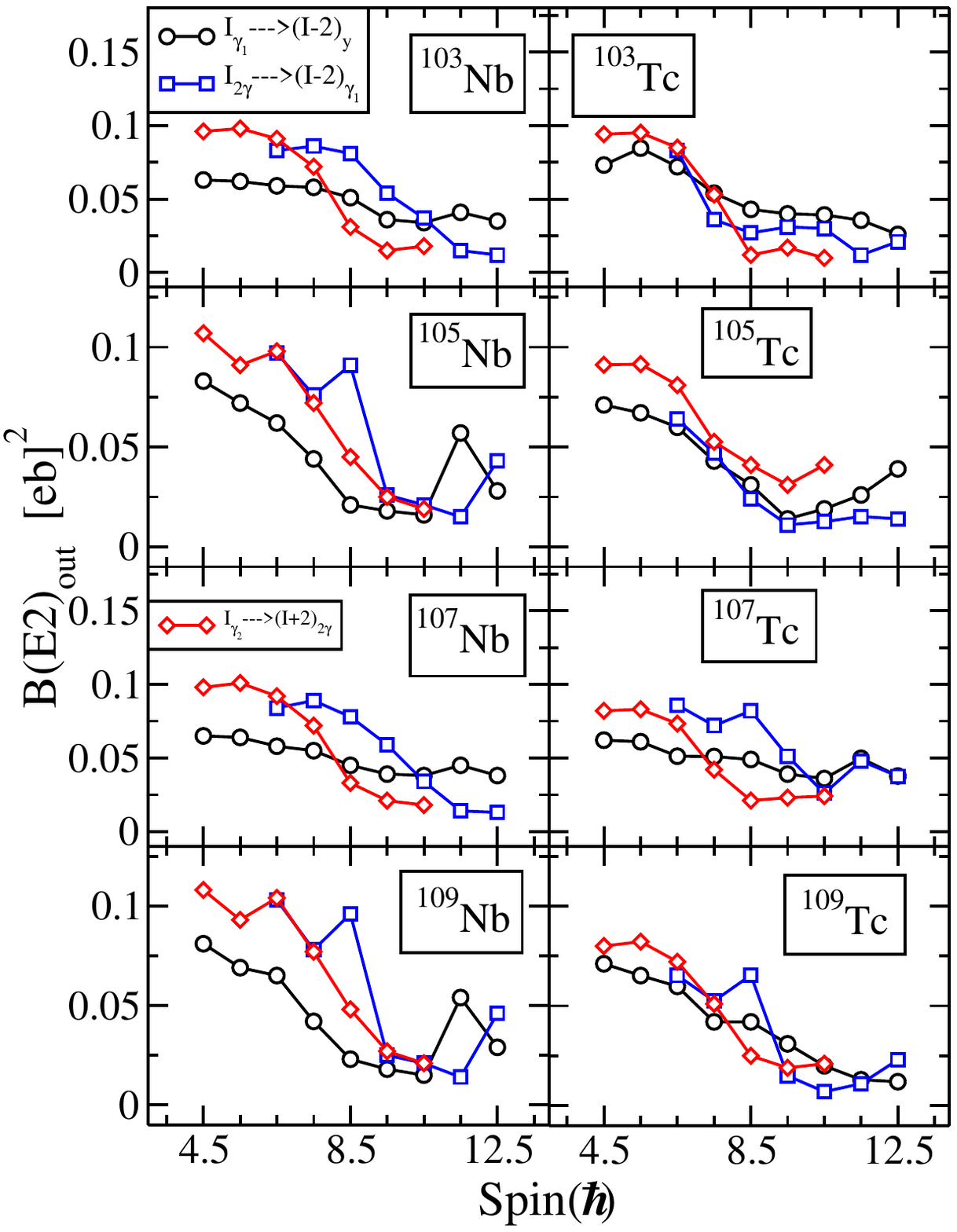} \caption{(Color online) $B(E2)$ transition probabilities ([eb]$^2$) from $\gamma_1$ band to yrast band (black curve), $2\gamma$ band to $\gamma_1$ band (blue curve) and $\gamma_2$ band to $2\gamma$ band (red curve) for $^{103,105,107,109}$Nb and $^{103,105,107,109}$Tc isotopes}. \label{BE2_out}
\end{center}
\end{figure}

For odd-mass nuclei, $^{103}$Nb was the first system where $\gamma$ and $2\gamma$ bands were identified \cite{WANG2009420}, and also
studied using the TPSM approach \cite{sh10}. In this mass region,
$\gamma$ band structures have now been delineated in several other odd-mass nuclei as well, and the purpose of the present work is  to
perform a systematic investigation of the high-spin band structures observed in isotopic chains of odd-mass Nb and Tc nuclides.
In particular, our focus will be on the fourth band observed in $^{103,105}$Nb nuclides, the nature of which has remained unexplored \cite{WANG2009420,Li054311}.
The first three bands have been classified as yrast, $\gamma$ and $2\gamma$ bands. However, it has been shown that
the fourth band observed cannot be characterized as $3\gamma$ band as the $B(E2)$ ratios obtained from the observed
transition intensities do not agree with the expected values \cite{Li054311}. It is demonstrated in the present work that this band
is another $\gamma$ band having the ``$K$'' value of $K_-=(K_0-2)$, where $K_0$ is  the $K$-value of the parent band. For the
$\gamma$ band, both $K_{\pm} = K_0 \pm 2$ values are possible. For the even-even system with $K_0=0$, $K=\pm 2$
are identical. However, for an odd-system with $K_0$
having a  half-integral value, the resultant two $K$ values are different with the possibility of having two $\gamma$ bands. 
It is demonstrated in the present study that the calculated third excited band in all the eight nuclides studied is the second
$\gamma$ band. 

The present manuscript is organized in the following manner. In the next section, a few details of the TPSM approach are provided. In section \ref{Sect_03},
the results for eight nuclides of Niobium and Technetium are presented and discussed. Finally, the present work is summarized
and concluded in section \ref{Sect_04}.

\section{Triaxial projected shell model approach}
\label{Sect.02}

The TPSM formalism is an amalgamation of mean-field and shell model approaches. In this approach, the triaxial Nilsson Hamiltonian is chosen as the 
mean-field and the pairing correlations are treated in the Bardeen-Cooper-Schrieffer approximation \cite{RS80}. The multi-quasiparticle configurations
are then constructed and projected onto good angular-momentum states using the projection method \cite{JAS21}. 
In the present study of odd-proton systems, the multi-quasiparticle configurations are composed of : one-proton, one-proton coupled to two-neutrons,
three-protons, and three-protons coupled to two-neutrons. The basis space is given by 
\begin{align}
&\hat{P}^{I}_{MK} \, a^{\dagger}_{\pi_1} |\Phi\rangle; \nonumber\\
&\hat{P}^{I}_{MK} \, a^{\dagger}_{\pi_1} a^{\dagger}_{\nu_1} a^{\dagger}_{\nu_2} |\Phi\rangle;\nonumber \\
&\hat{P}^{I}_{MK} \, a^{\dagger}_{\pi_1} a^{\dagger}_{\pi_2} a^{\dagger}_{\pi_3} |\Phi\rangle; \nonumber\\
&\hat{P}^{I}_{MK} \, a^{\dagger}_{\pi_1} a^{\dagger}_{\pi_2} a^{\dagger}_{\pi_3} a^{\dagger}_{\nu_1} a^{\dagger}_{\nu_2} |\Phi\rangle.
\label{eq:intrinsic_basis}
\end{align}
Here, \(|\Phi\rangle\) denotes the triaxially-deformed quasiparticle vacuum state. The operator \(\hat{P}^{I}_{MK}\) is the three-dimensional angular momentum projection operator \cite{RS80}, defined by
\begin{equation}
\hat{P}^{I}_{MK} = \frac{2I + 1}{8\pi^2} \int d\Omega \, D^{I}_{MK}(\Omega) \, \hat{R}(\Omega),
\label{eq:projection_operator}
\end{equation}
where the rotation operator \(\hat{R}(\Omega)\) is given by
\begin{equation}
\hat{R}(\Omega) = e^{-i\alpha \hat{J}_z} e^{-i\beta \hat{J}_y} e^{-i\gamma \hat{J}_z}.
\label{eq:rotation_operator}
\end{equation}
In the above expressions, \(\Omega = (\alpha, \beta, \gamma)\) represents the Euler angles (\(\alpha, \gamma \in [0, 2\pi]\), \(\beta \in [0, \pi]\)), and \(\hat{J}_z\), \(\hat{J}_y\) are the angular momentum operators. The Wigner D-function is denoted by \(D^{I}_{MK}(\Omega)\).

The rotational bands with the triaxial basis states (\ref{eq:intrinsic_basis}) are obtained by specifying different values for the $K$-quantum number in the projection operator (\ref{eq:projection_operator}). The three-dimensional angular-momentum
projection operator not only projects the angular-momentum quantum number, but also its projection along the z-axis \cite{LAMME1968492, LAMME1969609}.
The allowed values of the $K$-quantum
number for a given intrinsic state are obtained through the symmetry requirement \cite{KY95}.  
The projected basis space constructed from Eq.~(\ref{eq:intrinsic_basis}) is subsequently used to diagonalize a microscopic shell model Hamiltonian
composed of the spherical single-particle part and two-body interaction terms of quadrupole-quadrupole and pairing. These interaction terms are considered to
represent the most important correlations to describe the low-energy nuclear properties \cite{BarangerKumar}. The Hamiltonian is given by
\begin{equation}
\hat{H} = \hat{H}_{0} - \frac{1}{2} \chi \sum_{\mu} \hat{Q}^{\dagger}_{\mu} \hat{Q}_{\mu} - G_{M} \hat{P}^{\dagger} \hat{P} - G_{Q} \sum_{\mu} \hat{P}^{\dagger}_{\mu} \hat{P}_{\mu}.
\label{eq:total_hamiltonian}
\end{equation}
\(\hat{H}_{0}\) in the above equation represents the spherical part of the Nilsson potential \cite{Ni69} that includes a proper spin-orbit
interaction term. The strength parameter \(\chi\) in the quadrupole-quadrupole (QQ) interaction term is determined through the mean-field
self-consistent condition \cite{KY95}
\begin{equation}
\chi_{\tau\tau'} = \frac{\frac{2}{3}\epsilon \hbar\omega_\tau \hbar\omega_{\tau'}}{\hbar\omega_n \langle \hat{Q}_{0} \rangle_n + \hbar\omega_p \langle \hat{Q}_{0} \rangle_p}.
\label{eq:qq_strength}
\end{equation}
Here, \(\omega_\tau = \omega_0 a_\tau\), with \(\hbar\omega_0 = 41.4678 A^{-1/3}\) MeV. The
isospin-dependent factor \(a_\tau = [1 \pm (N-Z)/A]^{1/3}\) (where \(+\) applies to neutrons and \(-\) to protons) accounts for the asymmetry between neutron and proton potentials.
The monopole pairing interaction strength is obtained following the parameterization
\begin{equation}
G_{M} = \frac{G_1 \mp G_2 \frac{N-Z}{A}}{A},
\label{eq:monopole_pairing}
\end{equation}
where the minus (plus) sign corresponds to neutrons (protons).  Parameters \(G_1 = 20.25\) and \(G_2 = 16.20\) are typically adjusted to reproduce experimental odd-even mass differences, which serve as direct empirical measures of pairing correlations within nuclear systems.
The strength parameter \(G_Q\) is generally assumed to be proportional to the monopole pairing strength, typically following the relation \(G_Q \approx 0.16 G_M\).

The diagonalization of the Hamiltonian in Eq.~(\ref{eq:total_hamiltonian}) within the projected basis is performed using the Hill-Wheeler method \cite{RS80}.  This approach leads to the generalized eigenvalue equation 
\begin{equation}
\sum_{\kappa' K'} \left\{ \mathcal{H}_{\kappa K \kappa' K'} - E_\sigma \, \mathcal{N}_{\kappa K \kappa' K'} \right\} f^{\sigma}_{\kappa' K'} = 0,
\label{eq:hill_wheeler}
\end{equation}
where the Hamiltonian and norm kernels are defined as 
\begin{align}
\mathcal{H}_{\kappa K \kappa' K'} &= \langle \Phi_{\kappa} | \hat{H} \hat{P}^{I}_{K K'} | \Phi_{\kappa'} \rangle,\nonumber \\
\mathcal{N}_{\kappa K \kappa' K'} &= \langle \Phi_{\kappa} | \hat{P}^{I}_{K K'} | \Phi_{\kappa'} \rangle.
\label{ham_norm}
\end{align}
The wavefunctions obtained in the projection theory is given by

\begin{equation}
|\psi^{\sigma}_{IM}\rangle = \sum_{\kappa K} f^{\sigma}_{\kappa K} \, \hat{P}^{I}_{MK} |\Phi_{\kappa}\rangle.
\label{eq:hw_wavefunction}
\end{equation}
where $f^{\sigma}_{\kappa K}$ represents the expansion coefficients of the wavefunction in terms of the non-orthonormal projected basis states $\hat{P}^{I}_{MK} |\Phi_{\kappa}\rangle$. In the usual quantum-mechanical sense these coefficients are not probability amplitudes and in the present work, we have adopted the prescription of refs.~\cite{WANG2020,NaziraNP} to orthogonalize the wavefunction. In this procedure, the modified expansion coefficients $(g^{\sigma}_{\kappa K})$, which are expanded in terms of an orthonormal basis set in the following manner
\begin{equation}
g^{\sigma}_{\kappa K} = \sum_{\kappa^\prime K^\prime} f^\sigma_{\kappa^\prime K^\prime}\langle \kappa K| \hat{P}^I_{MK^\prime} | \Phi_{\kappa'} \rangle =\sum_{\kappa^\prime K^\prime} f^\sigma_{\kappa^\prime K^\prime}\mathcal{N}^{1/2 }_{\kappa K  \kappa^\prime  K^\prime}~,
\label{orthonormal_1} 
\end{equation}
where $|\kappa K \rangle $ is the orthonormal basis set and $\mathcal{N}$ is the norm matrix as defined in Eq.~(\ref{ham_norm}).


\begin{table}
\caption{Axial and triaxial quadrupole deformation parameters
$\epsilon$ and $\gamma$ (defined by $\gamma=\arctan(\epsilon' /
\epsilon$)) employed in the TPSM calculation for odd-mass Nb and Tc
isotopes. Axial deformations are taken from earlier works \cite{sh10,Li054311} and \cite{Moller1995}.}
\resizebox{\columnwidth}{!}
{\begin{tabular}{lllllllll}
\hline\hline        & $^{103}$Nb &$^{105}$Nb    &  $^{107}$Nb  &$^{109}$Nb &$^{103}$Tc &$^{105}$Tc     &  $^{107}$Tc  &$^{109}$Tc\\
\hline $\epsilon$ & 0.300 &  0.350 &    0.208    &  0.233   & 0.275    & 0.292     & 0.292& 0.242\\
$\epsilon'$       &0.16   & 0.13   &    0.100    &  0.110   & 0.14     & 0.13      & 0.135& 0.11 \\
       $\gamma$   &28$^o$     & 20$^o$    &    26$^o$         &  25$^o$       & 27$^o$        & 24$^o$         & 25$^o$ & 24$^o$  \\\hline\hline

\end{tabular}}\label{tab1}
\end{table}

\section{Results and discussion}
\label{Sect_03}

The TPSM calculations have been performed for the eight nuclides of $^{103,105,107,109}$Nb and $^{103,105,107,109}$Tc with the deformation parameters,
$\epsilon$ and $\epsilon'$, listed in Table \ref{tab1}. We would like to mention that some preliminary TPSM calculations have already been
performed for $^{103}$Nb \cite{sh10}. In this
section, we shall present the detailed TPSM results on the eight nuclides, including $^{103}$Nb, with a major focus on the fourth
excited band and the electromagnetic transition probabilities. In the previous communication, the properties of the fourth band
were not discussed and also the electromagnetic transitions were not reported.

We begin the presentation of the results with the depiction of the projected energies as a function of angular-momentum in Fig.~\ref{bd1}. In
this figure, we have chosen $^{105}$Nb nuclide as an illustrative example, and the results for other nuclides are quite similar. [These results
have been discussed in ref.~\cite{sh10} for $^{103}$Nb]. This
diagram, what is  referred
to as the band diagram, is quite useful as it provides information on the intrinsic structures of the bands before they are mixed in the final
diagonalization stage. It is noted from Fig.~\ref{bd1} that the ground-state band results from the angular-momentum projection of the intrinsic
one-proton state having quasiparticle energy of 0.98 MeV with $K = 5/2$. It needs to be emphasized that ``$K$'' value listed in Fig.~\ref{bd1} and in
other figures is  the projected ``$K$'' quantum number obtained from the angular-momentum projection formalism. The dominant
Nilsson configuration in the triaxial state having energy 0.98 MeV is $5/2^+[422]$ as known from the earlier works \cite{HOTCHKIS1991111, Luo_2005, Li054311}. 

It is evident from the symmetry requirement that the possible ``$K$'' values from a given parent configuration ($K_0$) are given
by $K=K_0 \pm 2, K_0 \pm 4, K_0 \pm 6,...$ \cite{KY95,sh10,Jeh21}. In the above case with $K_0 = 5/2$, the possible values
are $K = 9/2, 1/2, 13/2, 17/2,...$.
The bands with $K$ = 9/2, 13/2 and 17/2 are conventionally referred to as $\gamma$, 2$\gamma$ and 3$\gamma$ bands, respectively.
$\gamma$ and 2$\gamma$ bands have been observed 
in several odd-mass systems \cite{WANG2009420, sj3y_qnzp, Li054311,PhysRevC.74.054301, PhysRevC.74.024308, Luo_044310,Gu_054317, Gu_Long_2009}. However, the
second $\gamma$ band with $K = 1/2$, which results from the  $K = K_0-2$ combination, has not
been identified so far, and the main objective of the present work is to predict the excitation energy and other properties of this band
structure. In the following discussion, the second $\gamma$ band will be labelled as $\gamma_2$, and the normal $\gamma$ band as $\gamma_1$.

We would like to mention that $\gamma$ bands in odd-mass nuclei have also been investigated using the multi-phonon method (MPM)
\cite{PhysRevC.54.189} and detailed calculations have been performed for some representative nuclei
in  the $162 \le A \le 168$ region. It has been predicted using the MPM approach that $\gamma_2$ bands are lower in excitation energy as
compared to the $\gamma_1$ bands. However, in the mass $ \sim 100$ region, it is known that all the observed low-lying $\gamma$ bands
have the $\gamma_1$ structure \cite {Li054311, WANG2009420, PhysRevC.74.054301, Gu_Long_2009,Gu_Long_2010}. It is shown in the
present work that TPSM approach predicts the $\gamma_1$ band as the first excited band.

In Fig.~\ref{bd1}, the projected $K$ = 9/2, 13/2, 1/2 and 17/2 band structures from the intrinsic state with energy 0.98 MeV constitute the
lowest excited bands of the system.
What is interesting to note from the figure is that $\gamma_2$ band is lower in excitation energy
as compared to the 3$\gamma$ band. It will be demonstrated in the following that this band is the fourth band structure observed
in the $^{103}$Nb and $^{105}$Nb isotopes, the nature of which had remained unclear so far. In Fig.~\ref{bd1}, the band structures based on the single
quasiparticle state with energy of 1.64 MeV are also depicted, and it is noted that this $K = 1/2$ projected band exhibits a large signature staggering, and the
favoured signature of this band is lower in excitation energy as compared to the 3$\gamma$ band based on the ground-state configuration. It is
interesting to note that $\gamma_2$ band with $K = 1/2$, based on the $5/2[422]$ Nilsson configuration, shows no signature splitting
as it is a collective excitation. On the other hand, the excited $K = 1/2$ single-particle state has a predominant $1/2[400]$ Nilsson configuration and as
expected exhibits a large signature splitting. The projected bands from the three-quasiparticle configurations, displayed in
Fig.~\ref{bd1}, are situated at more than 2 MeV above the single quasiparticle states. It is noted from the figure that the projected band
from the three quasiparticle states with $K =7/2$ and quasiparticle energy of 2.75 MeV crosses the one-quasiparticle state at $I = 35/2$ and becomes
the yrast configuration. However, the experimental energies are  available only up to $I = 25/2$ for $^{105}$Nb, and this band crossing phenomenon cannot be
verified.

The TPSM excitation spectrum for $^{105}$Nb, obtained after the shell model diagonalization, is compared with the known experimental energies
in Fig.~\ref{105Nb}. Four band structures have been identified with the yrast band known up to $I$ = 25/2$^+$. The first and the second excited bands
have been categorized as $\gamma_1$ and 2$\gamma$ bands, respectively \cite{Li054311}. This
has been established from decay patterns of the two bands, and
from the similarity of the moments of inertia of these two bands with the yrast configuration. It is evident from
Fig.~\ref{105Nb} that TPSM calculations reproduce the energies of yrast, $\gamma_1$ and 2$\gamma$ bands quite well. The deviation for the highest
observed spin state of $I$ = 25/2$^+$ is 18 keV.

The nature of the third excited band with the tentative band head of $I$ = 9/2$^+$ and at an excitation energy of
1.687 MeV has remained unknown \cite{WANG2009420,Li054311}. It is expected from theoretical considerations that this band should correspond to
the 3$\gamma$ band structure. However, it has been shown that ratios of the $B(E2)$ transitions, deduced from the
transition intensity ratios of this band to the
2$\gamma$ band, are quite different from the corresponding ratios of the 2$\gamma$ band to the $\gamma_1$ band
\cite{Li054311}. These ratios should be similar for
the third excited band to be characterized as the 3$\gamma$ band. The TPSM calculations plotted in Fig.~\ref{105Nb} correspond to the lowest five
eigen-energies for each angular-momentum after diagonalization of the Hamiltonian. The corresponding wavefunctions of the five
eigenvalues for each angular-momentum are depicted in Fig.~\ref{wavefun}. It is evident from the figure that the yrast band has the
dominant $K = 5/2$ contribution from the quasiparticle
state with intrinsic energy of 0.98 MeV. The first and second excited bands have the dominant $K = 9/2$ and $K = 13/2$ configurations, respectively and are
projected states from the same quasiparticle state as that of the yrast state. The third excited band has the dominant $K = 1/2$ configuration and projected
from the same intrinsic
state as that of the lowest three eigenstates. This eigenstate results from the combination of $K_0-2$ and, therefor, corresponds to the second $\gamma$ band. The fourth calculated excited band
has the dominant $K = 17/2$ configuration and, consequently, corresponds to the 3$\gamma$ band structure. Considering that observed
fourth band in Fig.~\ref{105Nb} exhibits an excellent agreement with the projected $K = 1/2$ band, and also 3$\gamma$ band is predicted to be higher
in excitation energy, the observed fourth band is categorized as the $\gamma_2$ band. This band has also low-lying states of $I = 7/2, 5/2, 3/2$ and $1/2$, which
have not been populated in the experimental work.

To visualize the overall behaviour of the TPSM calculated band structures, the energies are subtracted by a
core contribution and are displayed in Figs.~\ref{eng_core_Nb} and \ref{eng_core_Tc} for the lowest five bands of Nb and Tc isotopes, respectively. 
The experimental energies are also provided in the figures for the known angular-momentum states. It is noted from the two figures that
yrast band for all the nuclides is reproduced quite well by the TPSM calculations. $\gamma$ bands are known for $^{103,105}$Nb and
$^{107,109}$Tc isotopes and it is evident from the comparison that TPSM approach provides a good description of the data. For the
2$\gamma$ band, a few states have been populated 
in $^{103}$Nb and $^{105}$Nb and it is again seen that TPSM calculations reproduce the energies of these states reasonably well.

Further, for the isotopes of $^{103,105}$Nb, a few states of the fourth band have also been populated and it has been stated earlier that the structure
of this band has remained unresolved \cite{WANG2009420,Li054311}. It is noted from Fig.~\ref{105Nb} that the energies of the known states of this band structure
agree well with the predicted energies of the $\gamma_2$ band. It is also observed from this figure that the predicted 3$\gamma$ band lies
at a higher excitation energy as compared
to the $\gamma_2$ band. As a matter of fact, it is noted from Figs.~\ref{eng_core_Nb} and \ref{eng_core_Tc} that $\gamma_2$ band is the third excited band for all the
eight nuclides studied.

It is apparent from the above discussion of the results that the lowest few band structures in all the eight nuclides studied belong to the
same intrinsic configuration. Although there is some mixing from other quasiparticle states, but this mixing is small in the low-spin region
as is evident from the wavefunction analysis. It is, therefore, expected that all the properties of the band  structures should be similar in the low-spin
region. These properties, for instance,  include alignments and the electromagnetic transition probabilities. We shall now turn our attention to the
discussion of these properties.

The calculated alignments from the TPSM and experimental quantities are depicted in Figs.~\ref{Nb_align} and \ref{Tc_align} for Nb and Tc isotopes,
respectively. The alignments have been calculated for the lowest five bands using the TPSM results and are shown on the right panels of the
two figures. The alignments evaluated from the experimental quantities are depicted on the left panels of the two figures. These are plotted for
four known bands in $^{103,105}$Nb and yrast bands for other nuclides. For $^{107,109}$Tc, alignments for the few known states of $\gamma$ bands
are also included in Fig.~\ref{Tc_align}. It is noted from Fig.~\ref{Nb_align} that both experimental and TPSM calculated alignments for $^{103,105}$Nb are
similar for all the band structures. In the TPSM calculated alignments for Nb isotopes, some deviation is noted in the high-spin region and is due to the
mixing from other quasiparticle states. In the Tc isotopes, deviations in the calculated alignments are larger as compared to the Nb isotopes and are due to the
stronger mixing of the quasiparticle configurations. In Fig.~\ref{wavefun103Tc}, the wavefunctions of the lowest five bands in $^{103}$Tc nucleus are depicted as an illustrative
example. It is noted from the figure that the three-quasiparticle crossing with the yrast ground state configuration occurs at a much
earlier spin value of $I = 25/2$ as compared to $^{105}$Nb. For
other bands, the mixing from three-quasiparticle states also becomes important at an earlier spin value. Due to this mixing, the calculated alignments for
Tc isotopes exhibit larger deviations for the five calculated band structures.

We have also evaluated the $B(E2)$ transitions for the eight Nb and Tc nuclides with the effective charges of $e_n=0.5e$ and $e_p=1.5e$. These transitions
from $\gamma_1$ to yrast, 2$\gamma$ to $\gamma_1$, and $\gamma_2$ to 2$\gamma$ bands are depicted in Fig.~\ref{BE2_out} and are also provided in Table \ref{BE2_Tab}.
It is noted from Fig.~\ref{BE2_out} that these inter-band transitions have similar behaviour as expected for bands belonging to the
same intrinsic state. Nevertheless, some deviations in $B(E2)$ transitions for different bands are evident from the figures and these are due to the
configuration mixing of the projected
states. The configuration mixing not only admits mixing from different intrinsic states, but also from different projected states belonging to the same intrinsic
state. This is evident from the wavefunction Figs.~\ref{wavefun} and \ref{wavefun103Tc}. The drop in the inter-band transition values with spin
is also due to the mixing of the configurations.

\begin{longtable*}{p{3.8cm}p{1.7cm}p{1.7cm}p{1.7cm}p{1.7cm}p{1.7cm}p{1.7cm}p{1.7cm}p{0.8cm}}
\caption{TPSM $B(E2)$ transition probabilities in (eb)$^2$ units for $^{103,105,107,109}$Nb and $^{103,105,107,109}$Tc isotopes.  }\\
 
\hline\hline
$I_i \rightarrow I_f$	&	$^{103}$Nb	&	$^{105}$Nb	&	$^{107}$Nb	&	$^{109}$Nb	&	$^{103}$Tc	&	$^{105}$Tc	&	$^{107}$Tc	&	$^{109}$Tc	\\
\hline\hline																	
\endfirsthead																	
\multicolumn{9}{c}{$I_{\gamma_1}\rightarrow  (I-2)_{y}$ }\\																	
\cline{1-9}																	
$9/2_{\gamma_1}\rightarrow 5/2_y$	&	0.063	&	0.083	&	0.065	&	0.081	&	0.073	&	0.071	&	0.062	&	0.071	\\
$11/2_{\gamma_1}\rightarrow 7/2_y$	&	0.062	&	0.072	&	0.064	&	0.069	&	0.085	&	0.067	&	0.061	&	0.065	\\
$13/2_{\gamma_1}\rightarrow 9/2_y$	&	0.059	&	0.062	&	0.058	&	0.065	&	0.072	&	0.059	&	0.051	&	0.059	\\
$15/2_{\gamma_1}\rightarrow 11/2_y$	&	0.058	&	0.044	&	0.055	&	0.042	&	0.054	&	0.043	&	0.051	&	0.042	\\
$17/2_{\gamma_1}\rightarrow 13/2_y$	&	0.051	&	0.021	&	0.045	&	0.023	&	0.043	&	0.031	&	0.049	&	0.042	\\
$19/2_{\gamma_1}\rightarrow 15/2_y$	&	0.036	&	0.018	&	0.039	&	0.018	&	0.041	&	0.014	&	0.039	&	0.031	\\
$21/2_{\gamma_1}\rightarrow 17/2_y$	&	0.034	&	0.016	&	0.038	&	0.015	&	0.039	&	0.019	&	0.036	&	0.021	\\
$23/2_{\gamma_1}\rightarrow 19/2_y$	&	0.041	&	0.057	&	0.045	&	0.054	&	0.036	&	0.026	&	0.049	&	0.013	\\
$25/2_{\gamma_1}\rightarrow 21/2_y$	&	0.035	&	0.028	&	0.038	&	0.029	&	0.026	&	0.039	&	0.037	&	0.012	\\
\hline																	
\multicolumn{9}{c}{$I_{2\gamma} \rightarrow  (I-2)_{\gamma_1}$ }\\																	
\cline{1-9}																	
$13/2_{2\gamma}\rightarrow 9/2_{\gamma_1}$	&	0.083	&	0.097	&	0.084	&	0.103	&	0.083	&	0.064	&	0.086	&	0.065	\\
$15/2_{2\gamma}\rightarrow 11/2_{\gamma_1}$	&	0.086	&	0.076	&	0.089	&	0.078	&	0.036	&	0.047	&	0.072	&	0.052	\\
$17/2_{2\gamma}\rightarrow 13/2_{\gamma_1}$	&	0.081	&	0.091	&	0.078	&	0.096	&	0.027	&	0.024	&	0.082	&	0.065	\\
$19/2_{2\gamma}\rightarrow 15/2_{\gamma_1}$	&	0.054	&	0.026	&	0.059	&	0.025	&	0.031	&	0.011	&	0.051	&	0.015	\\
$21/2_{2\gamma}\rightarrow 17/2_{\gamma_1}$	&	0.037	&	0.021	&	0.034	&	0.021	&	0.031	&	0.012	&	0.026	&	0.007	\\
$23/2_{2\gamma}\rightarrow 19/2_{\gamma_1}$	&	0.015	&	0.015	&	0.014	&	0.014	&	0.012	&	0.015	&	0.048	&	0.011	\\
$25/2_{2\gamma}\rightarrow 21/2_{\gamma_1}$	&	0.012	&	0.043	&	0.013	&	0.046	&	0.021	&	0.014	&	0.037	&	0.023	\\
\hline																	
\multicolumn{9}{c}{$I_{\gamma_2} \rightarrow  (I+2)_{2\gamma}$ }\\																	
\cline{1-9}																	
$9/2_{\gamma_2}\rightarrow 13/2_{2\gamma}$	&	0.096	&	0.107	&	0.098	&	0.108	&	0.094	&	0.091	&	0.082	&	0.081	\\
$11/2_{\gamma_2}\rightarrow 15/2_{2\gamma}$	&	0.098	&	0.091	&	0.101	&	0.093	&	0.095	&	0.095	&	0.083	&	0.082	\\
$13/2_{\gamma_2}\rightarrow 17/2_{2\gamma}$	&	0.091	&	0.098	&	0.092	&	0.104	&	0.085	&	0.081	&	0.073	&	0.072	\\
$15/2_{\gamma_2}\rightarrow 19/2_{2\gamma}$	&	0.072	&	0.072	&	0.072	&	0.077	&	0.053	&	0.052	&	0.042	&	0.051	\\
$17/2_{\gamma_2}\rightarrow 21/2_{2\gamma}$	&	0.031	&	0.045	&	0.033	&	0.048	&	0.012	&	0.041	&	0.021	&	0.025	\\
$19/2_{\gamma_2}\rightarrow 23/2_{2\gamma}$	&	0.015	&	0.025	&	0.021	&	0.027	&	0.017	&	0.031	&	0.023	&	0.019	\\
$21/2_{\gamma_2}\rightarrow 25/2_{2\gamma}$	&	0.018	&	0.019	&	0.018	&	0.021	&	0.011	&	0.041	&	0.024	&	0.021	\\

\hline\hline																

  \label{BE2_Tab}
\end{longtable*}

\section{Summary and conclusions}
\label{Sect_04}

In the present work, we have performed a detailed investigation of the high-spin band structures in odd-mass $^{103,105,107,109}$Nb and  $^{103,105,107,109}$Tc
isotopes with the primary focus on the $\gamma$ bands observed in these nuclei. In some of these nuclei, $\gamma$ bands have been observed, and also
in some cases 2$\gamma$ bands have been populated.  Further, a third excited band has been observed in $^{103,105}$Nb isotopes and the nature of this
band structure had remained unresolved.

We have employed the TPSM approach to provide an insight into the nature of the observed band structures for the eight nuclides. First of all, it has been demonstrated in
several earlier studies \cite{YK00,JY01,YS08,GH08,JG11,GH14,SJ18,JS21} that the TPSM approach provides a microscopic interpretation of the $\gamma$ vibrational bands observed in atomic nuclei.
The angular-momentum projection from a triaxial state results into several band structures having different values
of the ``$K$'' quantum number with  the permitted values of $K = K_0, K_0 \pm 2, K_0 \pm 4,K_0 \pm 6, ... $  \cite{KY95}.  
The band structure resulting from the $K_0+2$ combination with $K_0$ equal to zero for the ground-state of even-even systems is well established. This structure
is referred to as the $\gamma$ vibrational band observed in most of the even-even deformed and transitional nuclei. For odd-mass systems, $K_0+2$ and
$K_0-2$ have distinct values and can result into two different band structures.

In the mass $A \sim 100$ region, the first excited band observed is the $\gamma$ band resulting from the $K_0+2$ combination, and the second excited band is the
2$\gamma$ band, obtained from the $K_0+4$ combination. It has been demonstrated in the present work that the third excited band observed in $^{103,105}$Nb isotopes
is the second $\gamma$ band, resulting from the combination, $K_0-2$.  We have also studied the alignment and the electromagnetic transition probabilities, and it has
been shown that this band has similar properties as those of other three bands. Thus, indicating that the family of four band structures observed in $^{103,105}$Nb
belongs to the same triaxial intrinsic state.

For other studied nuclides, the high-spin data is quite limited and only in $^{107,109}$Tc isotopes, some states of the $\gamma$ band have been populated.
It is evident from the TPSM calculations of these nuclides that the predicted excitation energies and other properties are very similar
to those of $^{103,105}$Nb isotopes. We hope that in future experimental studies, it would be feasible to populate the high-spin band structures
in the studied Nb and Tc isotopes and validate the predictions of the present work, in particular, the presence of the second $\gamma$ configuration as the
third excited band.

\section*{Acknowledgements}
The authors are thankful to the Board of Research in Nuclear Sciences (BRNS), Department of Atomic Energy (DAE), Government of India for providing financial assistance under the Project No. 58/14/08/2025-BRNS/290.

\bibliographystyle{apsrev4-2}
\bibliography{bibliography}

\end{document}